\documentstyle[aps,preprint,psfig]{revtex}
\begin{document}
\draft
\preprint{{\tt Submitted to Surface Review and Letters, Jan.~30 1996}}
\title{Density functional theory study of 
Na at Al\,(111) and O at Ru\,(0001)}
\author{C. Stampfl}
\address{ Fritz-Haber-Institut der Max-Planck-Gesellschaft\\
Faradayweg 4-6, D-14\,195 Berlin-Dahlem, Germany}

\maketitle

\begin{abstract}
The success of density functional theory for the description of
the adsorption of atoms on surfaces is well established, and based on
recent calculations using gradient corrections, it has been shown that it
also describes well the dissociative adsorption of molecules at surfaces
-- admittedly however, the data base for reactions at surfaces is still
 somewhat small.
In the present paper the power of density functional
theory 
calculations is demonstrated by 
investigations for two different adsorption systems, namely,
one with a strongly electropositive adsorbate [Na on Al\,(111)]
and one with a strongly electronegative adsorbate [O on Ru\,(0001)].
In each case, new hitherto not expected
adsorbate phases have been predicted by the theory:
For Na on Al\,(111) the stability of a
``four-layer'' surface alloy was identified while for O on Ru\,(0001)
it was predicted that the formation of a $(1 \times 1)$-O adlayer 
should be possible which implies that the apparent saturation
coverage of $\Theta_{\rm O}=1/2$ is due to kinetic hindering.

\end{abstract}

\newpage
\section{Introduction}
 
It is of great interest to have an accurate theoretical
framework for the description of the interaction and reaction
of particles at surfaces. Calculations of this nature would give
insight and  understanding into the varied and complex processes
that take place between atoms, molecules, and solid surfaces.
In relation to this, information concerning atomic and electronic
structure, energy barriers, site specificity, bond formation
and scission, and the identification of
reaction pathways could be obtained.
Indeed, experimentally, a strong motivation for chemisorption 
studies with
well-defined single crystal surfaces is the
prospect of gaining deeper insight into the elementary  steps
governing heterogeneously catalyzed reactions.
Obviously a proper theoretical  description of
such processes at surfaces with accurate predictive power is
of enormous technological importance,
 particularly for example, in the field of catalysis.
In this respect, density
 functional theory is emerging as a very promising approach for
describing these phenomena which  involve the breaking and
 formation of chemical bonds, as for example,
 in the cases of dissociative adsorption, associative desorption, 
and
 chemical reactions. 

Many first-principles theoretical calculations have been performed 
for atoms,
molecules, and small clusters
 using quantum chemical methods
 \cite{cluster-rev}, and for bulk crystals, clean surfaces, and           
surfaces with adsorbates using methods originating from
solid state physics \cite{slab-rev}. Far fewer, however, have been 
performed
for systems containing both, that is, atoms and molecules at or  above
surfaces, as for example, gas--surface systems, or for
chemical reactions at surfaces.
Thus, schemes for
performing calculations for these kinds of systems typically employ
either a finite cluster of atoms or a supercell geometry, reflecting
the respective approaches of chemists and physicists.
The supercell method in particular, is proving to be
a very successful method for these kinds of  problems even though
its application to such systems  represents
a relatively new field.
See, for example,
the recent studies of hydrogen dissociation at various surfaces in 
Refs.~\cite{bird,hammer-cu111,hammer-reac,pehlke,kratzer,wilke2}.

From studies performed over the last few years, it has been demonstrated
that an improved description of the
exchange and correlation energy functional is necessary
for a realistic description of the physical
and chemical properties of certain systems.
The presently most successful and widely used
approach of achieving this is to use the generalized gradient
approximation (GGA).
Such an  approach results in a significantly improved description of free
atoms  and molecules, see for example, \cite{becke,fan,perdew,pople}.
Also for
the calculation of hydrogen dissociation,
where bonds are in the process of being broken
and reformed,
the GGA has been found to be of crucial importance for obtaining
 energy barriers in agreement with experiment.
Cohesive \cite{garcia} and chemisorption
energies obtained using the GGA
are also found to be considerably
closer to experiment than the LDA values,
for example, molecules at surfaces in which  there are strong charge
inhomogenuities in the system, e.g., for CO adsorption \cite{philipsen,hu1}.
We note that improvement brought about by using the GGA
is not limited to atoms and molecules in free space and
at surfaces,  it has also been found, for example, in a study for
different phases of ice to be necessary in order to describe the hydrogen
bond properly \cite{car}. Furthermore, it has been shown to
improve the coexistence pressure for the diamond structure to $\beta$-tin
phase transition in Si \cite{moll} and in describing a high-pressure
phase transition of SiO$_{2}$ \cite{hamann}.
There is still some difficulty in treating certain systems however,
as highlighted recently
in a test of twenty-four different local and gradient-corrected
DFT methods \cite{johnson}
for the energy barrier of the H + H$_{2} \longrightarrow
$H$_{2}$ + H reaction. In this study  it
was found that the energy barrier is significantly underestimated
when using the LDA {\em or} a GGA.
In this case, inclusion of a self-interaction correction
(SIC) was found to be important.

For surfaces, however, numerous calculations have demonstrated
that DFT within the LDA
represents a good description of the
quantum-mechanical many-body interactions
for well-bonded situations, where
calculated electron densities, atomic geometries, and total
energy differences are very reliable.
Nevertheless, the role,
of the GGA in calculations for adsorbates at surfaces, including
diffusion barriers,  is not yet clear.

In the present paper recent results
of  density functional
theory calculations for two quite different  
adsorption systems are discussed. The first system involves
a strongly electropositive adsorbate, namely,
sodium on Al\,(111) for  which the stability of an ordered
``four-layer'' surface alloy with a $(2 \times 2)$
periodicity was identified \cite{stampfl2}. This structure
was recently confirmed by  low energy electron diffraction (LEED) intensity
and surface extended x-ray adsorption fine structure (SEXAFS) 
studies \cite{burchhardt}. The second  system concerns
the  adsorption of oxygen on Ru\,(0001)
for which it was very recently 
predicted that the formation of a $(1 \times 1)$ adlayer should be 
possible which therefore implies that the apparent saturation
coverage of  $\Theta_{\rm O}=1/2$ observed under ultra high
vacuum conditions using molecular oxygen  is due only
to kinetic hindering  \cite{stampfl}.

\subsection{Calculation method}

The calculation method used to study the two adsorption systems is
the pseudopotential plane wave method
employing {\em ab initio},  fully separable
pseudopotentials where the adsorbate system is 
modelled using the supercell approach.
The calculation scheme \cite{neugebauer1,stumpf} affords
a fully self-consistent relaxation of the electrons and full
relaxation of the atomic positions using
damped molecular dynamics.
Calculations for Na on Al\,(111) were carried  out primarily
using the local-density
approximation (LDA) for the exchange-correlation functional
\cite{ceperley} where the non-linear form of the core-valence
exchange-correlation functional was taken into account
for the alkali metal atom \cite{neugebauer1,louie}. Further
details of the calculation can be found in
Ref.~\cite{stampfl2}.
One  additional  calculation
was  performed for the identified $(2 \times 2)$-Na/Al\,(111)
surface structure using the GGA
of Perdew {\em et al.}  \cite{perdew}.
In this respect,
we note that the GGA functional is used in creating the
pseudopotentials as well as in the surface calculations.
Thus the GGA is treated
in a consistent way which is
to be distinguished from
the inconsistent approach
where LDA pseudopotentials are used, with an otherwise GGA
treatment.
For the calculations for O on Ru\,(0001) we employed 
the GGA for all calculations because for our later, related work on the
catalytic oxidation of carbon monoxide reaction, use
of the GGA  is expected to be very important.
Details  of the calculations are given in Ref.~\cite{stampfl}.

\section{$(2 \times 2)$-Na/Al\,(111)}

Recent studies of Na and other alkali metal atoms on Al\,(111) 
have revealed that these systems contradict almost every expectation
of alkali metal adsorption behavior 
based on traditional understanding
(see, for example, Refs.~\cite{spotlight,bonzel,adams}).
In particular, at room temperature
where thermal energy allows activation barriers to be overcome,
evidence was presented from a combined surface extended x-ray adsorption
fine structure (SEXAFS) 
and density-functional theory (DFT) study
\cite{schmalz1} that an unexpected
 reaction takes place between the Na atoms
and the Al atoms at the surface: The Na atoms ``kick-out'' 
surface Al atoms and adsorb in their place, occupying
surface substitutional sites. The resulting structure has
a $(\sqrt{3} \times \sqrt{3})R30^{\circ}$  periodicity and
a local coverage of $\Theta=1/3$, where
$\Theta$ is the ratio of the number of
adsorbate atoms to the number of substrate atoms in an ideal layer.
The surface substitutional site for Na was later confirmed by a detailed
LEED intensity analysis \cite{nielsen11}.
In addition to sodium, it was subsequently shown 
that potassium \cite{stampfl1} and rubidium \cite{nielsen}
also adsorb substitutionally at room temperature with the same
periodicity and coverage.
Adsorption of lithium on Al\,(111) 
results in a $(\sqrt{3} \times \sqrt{3})R30^{\circ}$
structure and it has been proposed that
the Li atoms also
adsorb substitutionally \cite{nagao}; this, however,
has not been confirmed by a structural analysis.
Cesium, on the other hand,  forms a $(2\sqrt{3} \times 2\sqrt{3})R30^{\circ}$
structure at room temperature  and there is 
indication from high resolution core level spectroscopy
(HRCLS) studies \cite{jesper} that reconstruction of the
substrate also takes place, however, the exact atomic geometry  of
this phase is to date unknown.
At higher coverages ($\Theta=1/2$) of Na on Al\,(111), in contrast to
K, Rb, and Cs where $\Theta=1/3$ is the saturation coverage,
a ($2 \times 2$) structure  forms
in which there are two Na atoms per surface unit cell.
 Identification of the precise atomic
geometry of this phase had proved somewhat elusive: 
Early studies by Porteus \cite{porteus} and Hohlfeld and 
Horn  \cite{hohlfeld}
suggested models that involved unreconstructed Al substrates, namely,
in the
former study, it was proposed that there were three rotated domains
of $(2 \times 1)$ periodicity, 
and in the latter study,
the model was proposed to be comprised
 of two Na layers, each with a $(2 \times 2)$
periodicity.  
Based on HRCLS it was concluded \cite{jesper2} that these previous
models were incorrect and that a correct  model had to include strongly
intermixed Na/Al layers.
From a normal incidence 
standing x-ray wave-field (NISXW) study \cite{kerkar}
 a model was presented involving
two layers each of stoichiometry NaAl$_{2}$, while
from a recent scanning tunnelling microscopy (STM) study \cite{brune}, it 
was concluded  that in the surface unit cell
one Na atom  
occupies a substitutional site and the
other Na atom occupies a hollow site.
The, then unclear, situation regarding  knowledge of the detailed
atomic structure of  this $(2 \times 2)$ phase prompted  our theoretical
study \cite{stampfl2}. By performing calculations for a number
of possible surface geometries a particularly stable structure was
identified.
The validity of this structure,  
which constitutes a binary ``four-layer'' surface alloy,
was confirmed by
low energy electron diffraction (LEED)
and SEXAFS  \cite{burchhardt}.

\subsection{Atomic structure}

The atomic structure of $(2 \times 2)$-Na/Al\,(111) is shown in Fig.~1.
The Al layer containing atoms of the type Al$_{2}$
we refer to as the ``vacancy-layer''.  
The Na atoms, Na$_{s}$ and Na$_{f}$, 
occupy  substitutional  and fcc-hollow sites, respectively,  and 
in between these two Na layers, Al atoms, Al$_{1}$, are
positioned in hcp-hollow sites with respect to the ``vacancy-layer''. 
The structural parameters determined by 
DFT-LDA,  LEED, and SEXAFS are
given in Tab. I. It can be seen that the values agree very well;
the interlayer spacings obtained by all three methods deviate
by at most  by 0.13~\AA\,.
We note that the SEXAFS analysis could not 
uniquely determine the surface structure because good fits between
experimental and theoretical SEXAFS curves were 
also obtained for a different structural model, although it 
definitely excluded the NIXSW- and STM-derived models. 
The DFT-LDA and LEED results
exhibit, in particular, very good agreement:
they both identify the presence of small ($\approx$ 0.04 \AA\,)
lateral displacements of Al atoms in the ``vacancy-layer'' (Al$_{2}$)
and the layer beneath it, as well as a small rumpling  ($\approx$ 0.05 \AA\,)
in the  two Al layers   beneath the ``vacancy-layer''.

The calculations  for bulk Al using the GGA gave a 
lattice constant that is slightly 
larger than that obtained using the LDA: compare
 4.05~\AA\, to 3.98~\AA\,, i.e., it is  0.07~\AA\, or  1.83~\% 
larger.
These values can be compared to the experimental Al lattice constant
of 4.02~\AA\, \cite{chak} (at low temperature).
It is a general trend that GGA-determined lattice constants are
larger than the LDA-derived values \cite{garcia,filipi,kaxiras}.
The effect of the GGA with respect to the surface 
structure is found to yield Na-Al bond lengths somewhat longer
than  the  LDA result; the respective average
Na-Al bond lengths are 3.35~\AA\, and 3.26~\AA\,, i.e.
the value determined using the GGA is 2.7~\% longer than the result 
 obtained using the LDA. For comparison,  the corresponding value
 obtained from the LEED intensity analysis is 3.34~\AA\, which agrees
more closely with the GGA result.
The bond length of the Al atom, Al$_{1}$, to its nearest neighbors
in the ``vacancy layer'' however, increases by only 0.7~\% when
using the GGA as compared to the LDA result.
With respect to the above-mentioned lateral relaxations, and 
 the rumpling in Al layers five and six (i.e. the first and second
Al layers beneath the vacancy-layer, respectively), 
almost no change is found between the LDA and GGA results.
Similarly, the induced work function change is unaltered with a value
of $-$1.16~eV; the value and sign of which reflects
the electropositive nature of the Na atom.
We note that a similar effect was obtained in calculations for
the two $(\sqrt{3} \times \sqrt{3})$-Rb/Al\,(111) phases which
form at low temperature (on-top adsorption site)
and  room temperature (substitutional adsorption site) \cite{stampfl-unpub}.
 Here the bond lengths 
obtained for Rb in the on-top and substitutional sites using the LDA
were 3.30~\AA\, and 3.65~\AA\, respectively,
 and using the GGA the corresponding values were
3.45~\AA\, and 3.78~\AA\,.
 The  values obtained from a LEED  intensity analysis \cite{nielsen}
were 3.34~\AA\, and 3.72~\AA\, respectively, for the on-top and substitutional
sites. Similarly, as discussed below, the GGA results for the O-Ru
bond lengths are  slightly longer than those determined by LEED.
The tendency of a GGA to commonly give
slightly larger interatomic distances than experiment 
has been previously
noted \cite{perdew,kaxiras}, and it appears that this trend
extends to bond lengths of adsorbates at surfaces as well.

\subsection{Mass transport of substrate atoms}

The atomic structure of $(2 \times 2)$-Na/Al\,(111)  together with
that of  
$(\sqrt{3} \times \sqrt{3})R30^{\circ}$-Na/Al\,(111), for which Na
adsorbs substitutionally, and as mentioned above, 
likewise is formed for the K and Rb adsorbates,
indicates that a certain mass transport occurs on the surface as
explained in the following. Considering
first 
the $(\sqrt{3} \times \sqrt{3})R30^{\circ}$ structure,
 it is of interest to consider
how it forms and where the ``missing'' 
1/3 of a monolayer of Al atoms go.
It seems most natural to assume that
the alkali metal atoms ``kick-out'' Al atoms, i.e., a place
exchange occurs between the alkali metal
 atoms and Al atoms in the layer beneath.
In relation to this,
from DFT calculations \cite{neugebauer1,schmalz1}, 
a mechanism was identified which is
necessary for the kick-out reaction process to be significant,
namely, the re-bonding of the displaced substrate atoms
at kink sites at steps. 
This uses the fact that in thermal equilibrium kink
sites represent the thermodynamic reservoir for substrate
atoms and establishes that the chemical potential of the Al atoms equals
the cohesive energy.
As discussed in Ref.~\cite{adams}, 
it could also be possible, however, that
Al atoms desorb   {\em from} steps and diffuse 
to the regions between on-surface alkali metal 
atoms to form the substitutional structure, i.e., the on-surface
alkali metal atoms ``capture'' the Al atoms. Additionally it could be  
considered that at first
the alkali metal atoms kick-out Al atoms which then diffuse to
regions between on-surface alkali metal atoms, 
forming two terraces each with a substitutional structure.
Interestingly, from a HRCLS study of the temperature-dependent
 phase transition from the on-top to substitutional
$(\sqrt{3} \times \sqrt{3})R30^{\circ}$-Rb/Al\,(111) adsorption structures,
Al atoms  ``trapped'' between on-surface Rb atoms have
been identified \cite{lundgren}.

Some insight into the mechanisms involved in the formation of
the   $(\sqrt{3} \times \sqrt{3})R30^{\circ}$  substitutional
structures can be obtained by consideration of the activation
energy barrier  for the phase transition.
This activation energy barrier is related 
to the transition temperature through 
the Arrhenius equation.
For the kick-out mechanism, the transition state could be considered
as corresponding to a configuration involving, the alkali metal atom
plus an Al atom, over a vacancy in the Al\,(111) surface.
This transition state geometry will
yield a corresponding activation energy barrier 
(and transition temperature)
which {\em depends} on the species of alkali metal adsorbate since
the alkali metal  atom is involved in the geometry of the transition
state.
It may happen, however, that the barrier height is small or
comparable to that of an ``intermediate geometry''
corresponding to that in which the 
alkali metal atom is in the vacancy (adsorbed substitutionally) and
the  ejected Al atom in an isolated position on the  surface,
not yet re-bonded at a kink site at a step. In this case
the activation energy barrier
(and transition temperature) could be determined using this 
intermediate geometry.
This has been assumed in earlier work to estimate transition
temperatures \cite{adams,stampfl1}.
On the other hand, for the mechanism where we consider the possibility
that Al atoms desorb from steps
to regions {\em between} on-surface alkali metal atoms, the rate-determining
process is likely to be the desorption of Al atoms from the steps.
This mechanism will be largely {\em independent} of the species of the
on-surface alkali metal
adatoms and correspondingly so will the activation energy barrier and
transition temperature be.
Interestingly, from HRCLS \cite{jesper,lundgren}
 and second harmonic generation measurements \cite{wang},
it is found that the transition temperature
{\em does} depend on the adsorbate and is higher for
Rb  than for K, reflecting a higher
activation energy barrier for Rb: For Rb the transition
temperature was reported to be 
250 K \cite{lundgren} and for K temperatures
of 220 K \cite{jesper} and  210 K \cite{wang} were given.
 The fact that the transition temperature does depend
on the particular species of the alkali metal atom lends support to 
 the kick-out process. Additional support comes from
the STM study \cite{brune} in which small Al plateaus on the 
$(\sqrt{3} \times \sqrt{3})R30^{\circ}$   surface, not typically found
on the clean surface, were identified. 
Discussion along similar lines
has been presented in some detail in Ref.~\cite{adams}.

We now consider formation of the higher coverage 
 $(2 \times 2)$ phase.  The atomic composition
of this phase
reveals that another  interesting mechanism could be active:
On first consideration it would appear that no mass transport
of Al atoms is required in its creation,
that is,
the  ``missing'' Al atoms from the vacancy-layer are
seemingly re-bonded between the two Na layers (see Fig. 1).
However, because the $(2 \times 2)$ structure can be formed 
by adsorption of one-sixth of a 
monolayer of Na onto the  $(\sqrt{3} \times \sqrt{3})R30^{\circ}$
structure, in which every third surface Al atom is {\em missing},
a reverse process is implied, namely,
diffusion of one-third of a  monolayer of Al  
atoms back from kink sites at steps and reaction
with the  Na atoms to form the $(2 \times 2)$ phase.
A simplistic illustration of this process is provided in Fig. 2.
In the STM study
\cite{brune}  ``holes'' in terraces
 were observed to have formed  
on creation of the $(2 \times 2)$ phase.
The number of Al atoms taken from these ``holes'', however, 
was reported to correspond to only 1/12 of a monolayer, i.e., less
than the required 1/3 of a monolayer. This suggests that the additional
Al atoms come from steps or that they have undergone
long range mass transport and come from
areas on the surface not imaged.

We now turn to the second and quite different adsorption  system of 
O on Ru\,(0001).

\section{O on Ru\,(0001)}

 From recent experiments of the catalytic
oxidation of carbon monoxide performed at high gas partial pressures, 
evidence has been presented 
that Ru\,(0001) can support an unusually high concentration
of oxygen at the surface \cite{peden2,peden}. 
This is particularly interesting because it is well known that under
ultra high vacuum conditions (UHV) the saturation coverage is close
to $\Theta=1/2$ \cite{surnev} and very little in fact 
is actually known about possible
structures that may form under high O$_{2}$  pressures.
In order to investigate
the structure and stability of high coverage oxygen structures, we
performed density functional theory
calculations for various O-adlayers on Ru\,(0001) as well
as for clean Ru\,(0001), as reported in Ref.~\cite{stampfl}. 
The structures investigated were
the two ordered phases, $(2 \times 2)$ \cite{lindroos} and
$(2 \times 1)$ \cite{pfnur},
which form at room temperature
under UHV conditions
for coverages $\Theta=1/4$ and  $\Theta=1/2$, respectively,
and for several higher coverage
$(1 \times 1)$ structures with coverage $\Theta=1$.
The calculations for the $(2 \times 2)$ and $(2 \times 1)$ phases
provided a test of the accuracy of the calculations
through comparison with available LEED intensity analyses. From the 
comparisons it was found that very good
agreement with respect to the preferred adsorption site and the 
structural parameters was obtained. The investigation
for the higher coverage structures revealed 
that although a $(1 \times 1)$ phase is not
observed to form under UHV conditions using molecular oxygen, 
the adsorption of oxygen on Ru\,(0001) with a 
$(1 \times 1)$ structure 
is appreciably exothermic indicating that it should be able to form.
This furthermore implies that its formation at low O$_{2}$ pressures is 
prohibited by kinetic limitations due to activation energy barriers for
dissociation.

\subsection{The $(2 \times 2)$-O and $(2 \times 1)$-O phases on Ru\,(0001)}

For the $(2 \times 2)$ structure,
the theoretically obtained binding energy of O relative to the free O atom,
in the hcp-hollow site was found to be considerably more favorable than
in the fcc-hollow site where the energies are 5.55 eV
 and 5.12 eV, respectively.
 For the free O atom, we included the spin polarization energy
 which is calculated to be 1.521 eV \cite{ari}.
The DFT-GGA result is in accord with the LEED intensity determination
of the hcp-hollow adsorption site for oxygen in this phase.
The theoretical O-Ru   bond length is  2.10 \AA\, and
the first Ru-Ru interlayer spacing is found to be contracted by
2.7 \%
with respect to the bulk value
(using the centers of gravity of the first and second buckled Ru layers).
The corresponding LEED-derived values are  2.03 \AA\,   and
 2.1 \%, respectively.  As noted  above,
the calculated bond length  is somewhat longer
than the experimental one. Lateral and vertical relaxations of the
 first two Ru layers are
 induced by the O adatoms, the direction and magnitude of which
are well reproduced by the theory (see Ref.~\cite{stampfl} for
a comparison). Figure~3 shows the atomic geometry of the
$(2 \times 2)$-O/Ru\,(0001) structure as
determined by the calculations.
It can be seen that the three Ru atoms coordinated to the O 
adatom move laterally away from
it resulting in three Ru atoms in the first layer moving 
closer together; the fourth Ru atom does not move
laterally (it is prohibited by symmetry ) but it
is vertically displaced in towards
the bulk and slightly further 
than the other three Ru atoms. Similarly in the second Ru layer,
three Ru atoms move laterally towards each other and the fourth Ru atom,
below the O adatom, is vertically displaced in towards the bulk 
relative to the other three Ru atoms in the same layer.

Calculations for the  $(2 \times 1)$ structure show that
the obtained difference in
binding energy for oxygen in the 
fcc- and hcp-hollow sites is notably less than for that
of the lower coverage  $(2 \times 2)$ structure; the
binding energy of O in the hcp-hollow site is 5.28 eV and that for
the fcc-hollow site is 5.00 eV, a difference of 0.28~eV as
compared to 0.43~eV.
Again, adsorption in the hcp-hollow site is preferred,
 in agreement with the site
determined from the LEED intensity analysis \cite{pfnur}.
The oxygen-induced substrate relaxations
are particularly complex: the O atoms
atoms  adsorb in ``off'' hcp-hollow sites, i.e., they are displaced
slightly from the center of the site towards an on-top site,
and ``row-pairing'' and
buckling of the first two substrate layers occur. 
Figure~4 depicts the atomic structure
as obtained from the calculations.
On comparision
with the LEED determined structural parameters
 (see Ref.~\cite{stampfl}) we find that the
DFT-GGA calculations predicted all of these relaxations with
the one exception that our results indicated
row-pairing of the Ru atoms in the {\em second} Ru layer, as well as
in the first layer, and the LEED intensity analysis did not.
Both   DFT-GGA and LEED  find that the
 first two Ru-Ru interlayer spacings,
defined with respect to the
centers of gravity of the buckled atomic layers, correspond to the
bulk value to within 0.01~\AA\,.
The two theoretical  O-Ru bond lengths in the
$(2 \times 1)$ structure, each of approximately 2.08~\AA\,,
are very similar to that for the lower coverage
structure which was 2.10 \AA\,.
The value is again slightly  larger than that
obtained from the LEED intensity analysis, which for this structure, was
2.02~\AA\,. 

\subsection{The $(1 \times 1)$-O/Ru\,(0001) phase}

In view of the very good agreement of the results of
the DFT-GGA calculations and the LEED intensity analyses
 as outlined above,
 various higher coverage $(1 \times 1)$-O structures
with coverage  $\Theta=1$ were investigated.
In particular,
adlayers with O in the   fcc-, and hcp-hollow sites were
considered. Similarly to
the lower coverage structures, the hcp-hollow site was found to be
energetically more favorable
than the fcc-hollow site with respective
binding energies of 4.84~eV and 4.76~eV; the difference in binding energy
of 0.08~eV  between the two sites
again being reduced as compared to those of the
lower coverage structures. 
Calculations were also performed
with a higher energy cut-off and a larger {\bf k}-point
set, namely,  60 Ry and fourteen
special {\bf k}-points is the irreducible part of the Brillouin
zone.
The resulting binding energies differed by
only 0.03 eV and by 0.05~eV for the hcp- and fcc-hollow
sites, respectively,  and the resulting structural parameters differed by
not more than 0.02~\AA\, and by 0.04~\AA\, for the hcp- and fcc-hollow
sites, respectively, for the   two different basis sets.
These results are collected in Tab.~II.
In addition, using the larger basis set, the binding
energy of O in the on-top and bridge sites  was calculated.
These sites were found to be considerably less favorable that the
two hollow sites.
The obtained binding energies being 3.62, 3.93, 4.81, and 4.87~eV,
for the on-top, bridge, fcc- and hcp-hollow sites, respectively.
For  the hcp-hollow site,
the theoretical O-Ru bond length
of 2.03 \AA\, is slightly shorter than that of the lower coverage
structures
(compare 2.10~\AA\, and 2.08~\AA\, for the
$(2 \times 2)$ and $(2 \times 1)$ structures, respectively)
 and the first Ru-Ru interlayer spacing is found to be
{\em expanded} by $\sim$2.7~\% relative to the bulk spacing.
The surface atomic geometry is depicted in Fig.~5.

The value of 4.87 eV for the binding energy
of O in the hcp-hollow site on Ru\,(0001)
indicates that the $(1 \times 1)$ adlayer structure should be
able to form because
it is considerably larger  (by $\sim$ 1.8 eV)
than that which the O atoms have in O$_{2}$.
As noted above, under UHV conditions using O$_{2}$ the saturation
coverage is $\approx  0.5$ associated with the ordered
$(2 \times 1)$ superstructure.
Thus, the reason that a $(1 \times 1 )$ adlayer with coverage
$\Theta=1$ does not form under UHV conditions
is apparently due to a kinetic hindering for  the
dissociation of O$_{2}$.  It is probable that
such a hindering
may be overcome  at high O$_{2}$  pressures and elevated temperatures
as used in the catalytic reactor experiments \cite{peden2,peden}
due to the high attempt frequency. Alternatively,
if {\em atomic} oxygen is offered, the $(1 \times 1)$-O structure may be
realized as we predicted in Ref.~\cite{stampfl}.

The coverage dependence of the binding energy
described above,
in which it becomes less  favorable with
increasing coverage,  reflects a repulsive
interaction between the adsorbates and implies
that no island formation is expected to occur in the coverage regime
of $\Theta=1/4$ to $\Theta=1$.
Concomitantly, the difference in binding energy
between the fcc- and hcp-hollow sites becomes less with 
increasing coverage.
In fact, for the full monolayer this difference is quite small. 
However, because
the full monolayer is reached successively via the other (lower coverage)
phases, for which the hcp-hollow
site is clearly favored, we expect a nearly perfect
hcp-hollow site occupation, i.e. only  few fcc-hollow 
site dislocation structures, for the $(1 \times 1)$ oxygen layer.
In the coverage range investigated,  
the work function increases with increasing coverage 
in accordance with the high electronegativity of oxygen
and in good agreement with experiment \cite{surnev}.
For the $(1 \times 1)$-O structure, the value of the work function 
change is about 1.11~eV.

Because the $(1 \times 1)$-O adlayer is appreciably exothermic it 
prompts the consideration of whether an even higher 
coverage of oxygen can adsorb at the surface.
To investigate this possibility
we performed calculations  for three structures 
with an O coverage  of $\Theta=1.25$.
These structures are shown in Fig.~6.
In the first structure, depicted  in Fig.~6a,
there are four O atoms in hcp-hollow sites in the
$(2 \times 2)$ surface unit cell and one in an fcc-hollow site.
This structure was found to be unstable in that the  oxygen atom
in the fcc-hollow site did not form  a bond with the substrate and
lifts off the surface.
In the second geometry considered, illustrated in Fig.~6b,
an O$_{2}$ dimer is adsorbed in the hcp-hollow
site of the surface unit cell with the molecular axis
perpendicular to the surface. In the remaining three hcp-hollow sites,
O atoms are positioned.
Likewise, this structure was also found to be unstable.
These results indicate that
the optimum coverage chemisorbed {\em on the surface} is $\Theta$=1.
In the last structure considered, as shown in Fig.~6c, there are
four O atoms in hcp-hollow sites in the surface unit cell  and 
one O atom in an
octahedral (subsurface) site between the first and second
Ru layers. This structure was found to be metastable where
the average adsorption energy
per O atom in the surface unit cell
is 4.09~eV; alternatively stated, a free O atom gains 1.07~eV if it can adsorb
in the subsurface octahedral site of the O-covered surface.
This geometry is, however,  less favorable (by an average of 0.37~eV per atom)
than that with
four O atoms in hcp-hollow sites 
and the fifth
O atom in an O$_{2}$ dimer in gas-phase.
That is, while O adsorption  under the
O-covered surface in an octahedral site 
 is metastable
with respect to free O atoms, if O$_{2}$ is used,
it is energetically more favorable to retain the  $(1 \times 1)$
adlayer structure with the additional O atom in the O$_{2}$ dimer in the
gas-phase. If, however, atomic oxygen is used it is possible that the oxygen
atoms may enter the subsurface region, provided they can 
overcome the diffusion energy barrier for passing through the
first Ru layer.

\section{conclusion}

Using density functional theory calculations
it has been possible to identify new  and unexpected
adsorbate phases, as well as obtaining insight
into the behavior of the adsorption systems
in general. 
In the present paper  this was demonstrated for
the $(2 \times 2)$-Na phase on Al\,(111) which 
represents a
binary ``four-layer'' surface alloy that is able to form
even though the constituent atoms are
immiscible in bulk, and for the adsorption of oxygen on Ru\,(0001)
for which it was found that 
a $(1 \times 1)$-O adlayer structure 
should be able to form.
As indicated in the introduction, the 
application of similar theoretical approaches
to those mentioned in the present paper, to studies into the  interaction
and chemical reaction of  atoms and molecules  at surfaces
is expected to be a particularly active area due
to the relevance of  these processes in heterogeneous
catalysis.

\vspace{1cm}
{\bf Acknowledgements}

Valuable  discussions with M. Scheffler and K. Kambe are gratefully
acknowledged.

\begin{table}
\begin{tabular}{l|cccccc}
Interlayer spacing (\AA) & Z(Na$_{f}$-Al$_{1}$) &
 Z(Al$_{1}$-Na$_{s}$)
& Z(Na$_{s}$-Al$_{2}$)
& Z(Al$_{2}$-Al$_{3}$)  &
Z(Al$_{3}$-bulk)\\
\hline
LEED	&  0.85 & 0.55 & 1.52 & 2.25 & 2.28 \\
SEXAFS  &  0.75 & 0.70 & 1.50 &  \\
DFT-LDA &  0.72 & 0.62 & 1.46 & 2.23 & 2.27 \\
\hline
DFT-GGA &  0.86 & 0.49 & 1.63 & 2.29 & 2.33 \\
\end{tabular}
\caption{Interlayer spacings, Z, between the atomic
layers indicated in parenthesis
(see Fig.~1).
For the rumpled Al layers, five and six, the interlayer distances are given 
with respect to the center of gravity of the respective  layer.
In relation to this, we note that in Refs. 
\protect\cite{stampfl2,burchhardt}, 
the interlayer spacings for the DFT-LDA results
were given with respect to the positions of the three  Al atoms per unit
cell which have the same vertical position and not
with respect to the center of gravity of the rumpled layer.}
\end{table}

\newpage
\begin{table}
\begin{tabular}{cccccc|c}
 & \multicolumn{6}{c}{$(1 \times 1)$-O/Ru\,(0001) hcp-hollow site}  \\
\hline
 & O-Ru &  $d_{z,1}$
& $d_{z,2}$ & $d_{z,3}$  & $d_{z,{\rm bulk}}$ & E$_{b}$\\
\hline
DFT-GGA (40 Ry) & 2.04 &  1.28 & 2.27 & 2.19 & 2.19 & 4.84 \\
DFT-GGA (60 Ry) & 2.03 &  1.26 & 2.24 & 2.17 & 2.19 & 4.87\\
\hline
 & \multicolumn{6}{c}{$(1 \times 1)$-O/Ru\,(0001) fcc-hollow site}  \\
\hline
DFT-GGA (40 Ry) & 2.05 &  1.29 & 2.33 & 2.13 & 2.19 & 4.76 \\
DFT-GGA (60 Ry) & 2.03 &  1.27 & 2.29 & 2.13 & 2.19 & 4.81\\
\end{tabular}
\caption{Structural parameters for $(1 \times 1)$-O/Ru\,(0001)
with O in the hcp- and fcc-hollow sites for the different basis sets
(see text).
O-Ru, $d_{z}$, and E$_{b}$ represent, the O-Ru bond length,
 the interlayer spacings (in \AA ngstrom), and
binding energy (in eV), respectively.
}
\end{table}

\begin{figure}
\psfig{figure=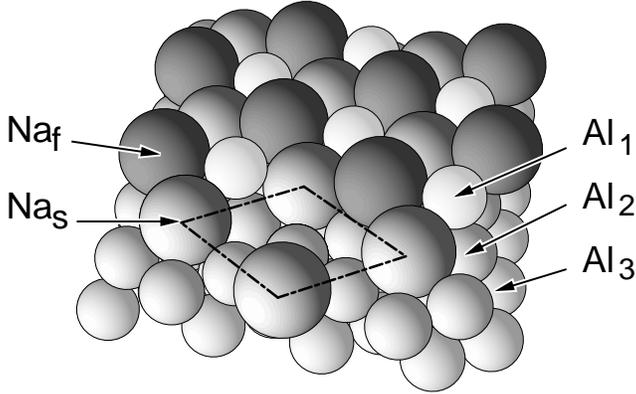,width=10cm}
\caption{The atomic structure of $(2 \times 2)$-Na/Al\,(111). Some atoms
have been removed to expose the lower layers where the surface unit
cell is also drawn. Na$_{f}$ and Na$_{s}$ represent, respectively, the
Na atoms in fcc-hollow and substitutional sites and Al$_{1}$ the
Al atoms located in hcp-hollow sites, with respect to the 
vacancy-layer. Al$_{2}$ and Al$_{3}$ represent the Al atoms in the
vacancy-layer and in the layer below, respectively.}
\end{figure}

\begin{figure}
\psfig{figure=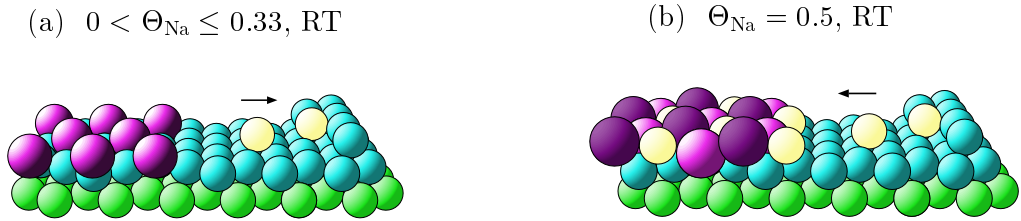,width=15cm}
\caption{Possible mass transport mechanism.
(a) Island formation occurs \protect\cite{brune} with a 
$(\protect\sqrt{3} \protect\times \protect\sqrt{3})R30^{\protect\circ}$
structure with Na atoms in substitutional sites, for coverages
$0 < \Theta_{\rm Na} \protect\leq  1/3$: Every third Al atom is
``missing''; Al atoms diffuse to steps. (b) Surface alloy formation
with a $(2 \times 2)$ geometry implies diffusion of Al atoms back from steps. 
}
\end{figure}

\begin{figure}
\psfig{figure=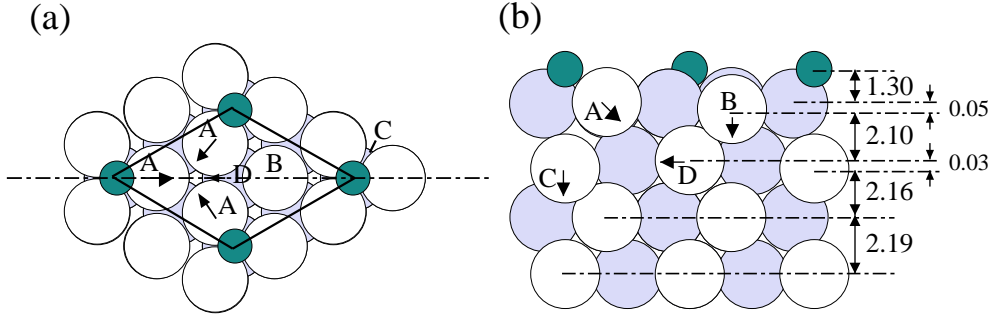,width=15cm}
\caption{Top view (a) and side view (b) of the atomic
 geometry of $(2 \times 2)$-O/Ru\,(0001).
The arrows (not drawn to scale)
 indicate the direction of the displacements of the substrate atoms
with respect to the bulk positions.
The dashed line  in (a) indicates the plane of the cross-section used
in (b).  Small dark grey circles represent
oxygen atoms and large white and grey circles represent Ru atoms,
where the latter correspond to those lying in the next plane.
Interlayer spacings are given in \AA ngstrom.}
\end{figure}

\begin{figure}
\psfig{figure=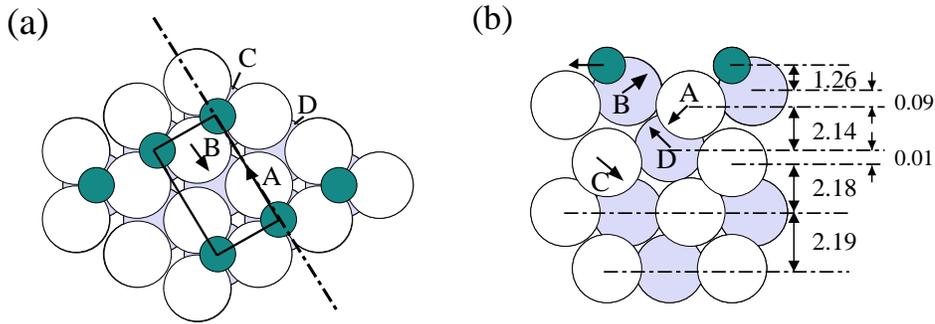,width=15cm}
\caption{Top view (a) and side view (b) of the atomic
 geometry of $(2 \times 1)$-O/Ru\,(0001).
The arrows (not drawn to scale)
indicate the direction of the atomic displacements.
The dashed line in (a) indicates the plane of the cross-section used
in (b).
 Small dark grey circles represent
oxygen atoms and large white and grey circles represent Ru atoms,
where the latter correspond to those lying in the next plane.
Interlayer spacings are given in \AA ngstrom.}
\end{figure}

\begin{figure}
\psfig{figure=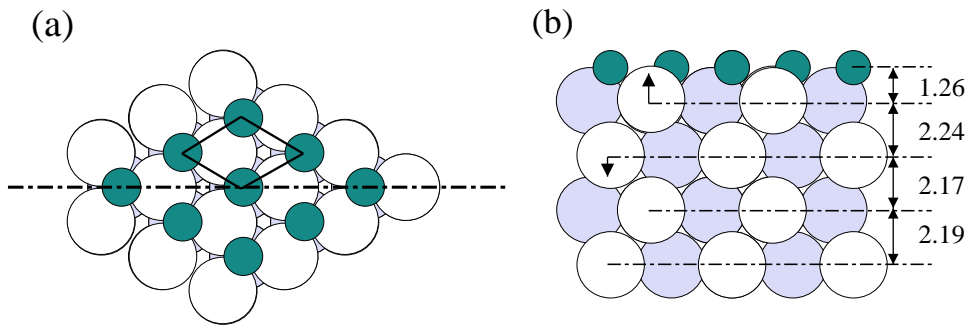,width=15cm}
\caption{Top view (a) and side view (b) of the atomic
 geometry of $(1 \times 1)$-O/Ru\,(0001) with O in the hcp-hollow
site (obtained with a 60 Ry cut-off and fourteen special
{\bf k}-points in the irreducible part of the Brillouin zone).
The arrows (not drawn to scale) indicate the direction
of the displacements of the substrate atoms
with respect to the bulk positions.
 Small dark grey circles represent
oxygen atoms and large white and grey circles represent Ru atoms,
where the latter correspond to those lying in the next plane.
Interlayer spacings are given in \AA ngstrom.}
\end{figure}

\begin{figure}
\psfig{figure=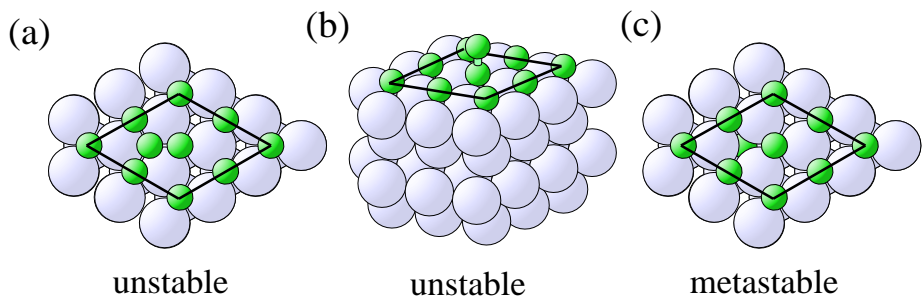,width=15cm}
\caption{Structures considered for O on Ru\,(0001) for  a coverage
of $\Theta=1.25$. The Ru atoms are depicted as the large
circles and the O atoms as the small circles. The surface unit
cell is indicated.}
\end{figure}

\end{document}